\documentclass[12pt]{article} 
\usepackage{a4}
\usepackage{amsmath}
\usepackage{amssymb}
\usepackage{latexsym}
\usepackage{amssymb}
\usepackage{epsfig}
\usepackage{natbib}
\def \ii{{\mathrm{i}}}

\def \d{{\mathrm{d}}}

\def \pd{\partial}

\def \e{{\mathrm{e}}}

\def \BF{\boldsymbol{F}}

\def \Bx{{\boldsymbol{x}}}
\def \By{{\boldsymbol{y}}}

\begin{document}
\title{{\bf 
On non-singular crack fields 
in Helmholtz type enriched elasticity theories
}}
\author{
Markus Lazar~$^\text{a,}$\footnote{
{\it E-mail address:} lazar@fkp.tu-darmstadt.de.}
\ 
and Demosthenes Polyzos~$^\text{b,}$\footnote{
{\it E-mail address:} polyzos@mech.upatras.gr.}
\\ \\
${}^\text{a}$ 
        Heisenberg Research Group,\\
        Department of Physics,\\
        Darmstadt University of Technology,\\
        Hochschulstr. 6,\\      
        D-64289 Darmstadt, Germany\\
\\
${}^\text{b}$ 
Department of Mechanical and Aeronautical Engineering\\
University of Patras\\
GR-26500 Patras, Greece
}

\date{\today}    
\maketitle

\begin{abstract}
Recently, simple non-singular stress fields of cracks of mode I and mode III
have been published by~\citet{Aifantis09,Aifantis12,Isaksson13} and ~\citet{Isaksson12}. 
In this work we investigate the physical meaning and interpretation 
of those solutions 
and if they satisfy important physical conditions (equilibrium, boundary 
and compatibility conditions). 
\\

\end{abstract}
\section{Introduction}
During the last years, some non-singular crack fields have been published in 
the literature~\citep{Aifantis09,Aifantis12,Isaksson13,Isaksson12} 
neglecting equilibrium, boundary and compatibility conditions.
The aim of that research was the regularization of the classical singular crack
fields. In fact, the non-singular crack fields are zero at the crack tip.
However, not any equilibrium condition was used
by~\citet{Aifantis12,Isaksson13} and \citet{Isaksson12}.
Therefore, it is  doubtful if their results are correct from the
point of view of fracture mechanics.

On the other hand, \citet{AE83,Eringen83} and \citet{Eringen84} 
(see also~\citet{Eringen02}) 
investigated cracks in the framework of nonlocal elasticity of Helmholtz type 
in the 80s. 
\citet{Eringen84,Eringen02} found a non-singular stress of a mode III crack 
zero at the crack tip. For the mode I crack problem, 
using appropriate boundary conditions, \citet{AE83} (see also \citet{Eringen02})
found a non-singular stress finite at the crack tip and becoming zero
inside the crack. A regularization procedure was also discussed for 
crack curving in~\citet{Eringen83}.

The aim of this paper is to show that  the 
recent crack solutions given 
by~\citet{Aifantis09,Aifantis12,Isaksson12,Isaksson13}
cannot be the correct solutions of a nonlocal or gradient compatible elastic fracture
mechanics problem. 
Therefore, the modest goal of the present paper 
is not to give new solutions
but rather to discuss existing and recently given crack solutions using
gradient enhanced elasticity theories.

The paper is organized as follows:
section~2 provides the basics of the theories of nonlocal elasticity and 
strain gradient elasticity. 
Next, section~3 explains why the non-singular mode III crack solution
given by~\citet{Aifantis09,Aifantis12} cannot be considered as a solution of a
nonlocal and gradient elastic fracture mechanics problem. 
The same is explained in section~4 for the mode I crack solution 
given by~\citet{Aifantis12,Isaksson12} and \citet{Isaksson13}. 
Finally, in section~5 a possible way-out for the physical interpretation of
the non-singular solutions is discussed.

\section{Theoretical framework}
In this section we outline 
the basics of the theories of nonlocal elasticity and gradient elasticity.

\subsection{Theory of nonlocal elasticity of Helmholtz type}
In the theory of nonlocal elasticity (e.g.,~\citet{Eringen02,Eringen83b}), 
the so-called nonlocal stress tensor $t_{ij}$ is defined at any point
$\Bx$ of the analyzed domain of volume $V$ as
\begin{align}
\label{HL-int}
t_{ij}(\Bx)&=\int_V\alpha(|\Bx-\By|) \sigma^0_{ij}(\By)\,\d V(\By)\,,
\end{align}
where $\alpha(|\Bx-\By|)$ is a nonlocal kernel
and $\sigma^0_{ij}$ is the stress tensor of classical elasticity defined at the point $\By\in V$ as
\begin{align}
\label{HL}
\sigma^0_{ij}(\By)&=\lambda\, \delta_{ij} e^0_{kk}(\By)+2\mu e^0_{ij}(\By)
\end{align}
with $\lambda$, $\mu$ are the Lam\'e constants, $\delta_{ij}$ is the 
Kronecker delta and $e^0_{ij}$ denotes the classical strain tensor,
which is the symmetric part of the classical distortion tensor
\begin{align}
e^0_{ij}&=\frac{1}{2}\big(\beta^0_{ij}+\beta^0_{ji}\big)\,.
\end{align}
We employ a comma to indicate partial derivative with respect to rectangular 
coordinates $x_j$, i.e. 
$t_{ij,j}=\frac{\pd t_{ij}}{\pd x_j}$.
As usual, repeated indices indicate summation.

In absence of body forces, the nonlocal stress tensor satisfies the 
equilibrium condition
\begin{align}
\label{EC-NL}
t_{ij,j}=0\,,
\end{align}
which means that the stress is self-equilibrated.
In addition, the classical stress tensor fulfills the equilibrium equation of 
classical elasticity
\begin{align}
\label{EC-cl}
\sigma^0_{ij,j}=0\,.
\end{align}

If the nonlocal kernel function $\alpha(|\Bx-\By|)$ is the Green function
(fundamental solution) of the differential operator
$L=1-\ell^2\Delta$, i.e.
\begin{align}
\label{a-HE}
(1-\ell^2\Delta) \alpha(|\Bx-\By|)=\delta(\Bx-\By)
\end{align}
with $\ell$, $\Delta$, $\delta$ being 
a characteristic length scale ($\ell\ge 0$),
the Laplacian and the Dirac delta function, respectively, 
then the integral relation~(\ref{HL-int}) reduces to 
the inhomogeneous Helmholtz equation
\begin{align}
\label{t-HE}
(1-\ell^2\Delta) t_{ij}=\sigma^0_{ij},
\end{align}
where the classical stress is the source for the nonlocal stress.
The natural boundary condition reads
\begin{align}
\label{BC-NE}
t_{ij}n_j=\hat{t}_i\,,
\end{align}
where $n_i$ and $\hat{t}_i$ represent the normal to the external boundary and 
the prescribed boundary tractions, respectively.
In nonlocal elasticity, no nonlocal strain exists.
Thus, using a nonlocal kernel, being a Green function, yields 
a differential equation for $t_{ij}$ instead of an integrodifferential equation
in the `strongly' nonlocal theory
with seemingly and physically equivalent solution at the output.
In such a `weakly' nonlocal elasticity the concept of a gradient theory
might be used~\citep{Maugin,Maugin2}.
The `weakly' nonlocal theory of elasticity represented 
by Eqs.~(\ref{EC-NL})--(\ref{t-HE}) is called of 
Helmholtz type because the Helmholtz operator, $L=1-\ell^2\Delta$, enters
in the form of equations~(\ref{a-HE}) and (\ref{t-HE}).

It was pointed out by~\citet{Eringen83} that 
the stress field $t_{ij}$ of a crack is obtained by solving Eq.~(\ref{t-HE}), 
subject to regularity conditions, i.e., $t_{ij}$ must be bounded 
at the crack tip and at infinity. This is borne out from the problems
of non-singular dislocations~\citep{Eringen02}. 
At large distance from the crack tip, the classical solution will
approximate the stress field well, namely if $\ell\to 0$,  
Eq.~(\ref{t-HE}) gives $t_{ij}\to\sigma^0_{ij}$.
This also suggests that one may obtain a full solution of Eq.~(\ref{t-HE})
and match it to the outer solution $\sigma^0_{ij}$ in order to obtain 
a non-singular solution.

\subsection{Theory of gradient elasticity of Helmholtz type}
In the theory of gradient elasticity (see, e.g., \citet{Mindlin,ME,Eshel,Jaunzemis}), 
the equilibrium condition is given by
\begin{align}
\label{EC-GE}
\tau_{ij,j} -\tau_{ijk,jk}=0\,,
\end{align}
where $\tau_{ij}$ is the Cauchy-like stress tensor and 
$\tau_{ijk}$ is the so-called double-stress tensor.
It can be seen in Eq.~(\ref{EC-GE}) that the Cauchy-like stress tensor 
$\tau_{ij}$ is, in general, not self-equilibrated.
The natural boundary conditions in strain gradient elasticity 
are much more complicated than the corresponding ones in nonlocal elasticity; 
they read (see, e.g.,~\citet{ME,Jaunzemis})
\begin{align}
\label{BC-GE}
\left.
\begin{array}{r}
\displaystyle{\big(\tau_{ij}-\pd_k\tau_{ijk}\big)n_j-\pd_j\big(\tau_{ijk} n_k\big)
+n_j\pd_l\big(\tau_{ijk} n_k n_l\big)=\bar{t}_i}\\
\displaystyle{\tau_{ijk}n_jn_k=\bar{q}_i}\\
\end{array}
\right\}
\qquad\text{on}\qquad \pd\Omega\,,
\end{align}
where $\bar{t}_i$ and $\bar{q}_i$ are the prescribed Cauchy traction vector and the 
prescribed double stress traction vector, respectively. 
Moreover, $\pd\Omega$ is the smooth boundary surface of the domain $\Omega$ 
occupied by the body.

In a simplified version of strain gradient elasticity, called gradient 
elasticity of Helmholtz type (e.g.,~\citet{LM05,Lazar13,Polyzos}), 
the double stress tensor is nothing but the gradient of the Cauchy-like 
stress tensor multiplied by $\ell^2$
\begin{align}
\label{CE-GE}
\tau_{ijk}=\ell^2\tau_{ij,k}\,,
\end{align}
and the Cauchy-like  stress tensor reads
\begin{align}
\label{HL2}
\tau_{ij}=C_{ijlk} \beta_{kl}\,,
\end{align}
where $\beta_{ij}$ denotes the elastic distortion tensor and 
$C_{ijkl}$ is the tensor of the elastic moduli given by
\begin{align}
\label{C}
C_{ijkl}=\mu\big(
\delta_{ik}\delta_{jl}+\delta_{il}\delta_{jk}\big)
+\lambda\, \delta_{ij}\delta_{kl}\,.
\end{align} 
Substituting Eq.~(\ref{CE-GE}) into (\ref{EC-GE}), Eq.~(\ref{EC-GE})
simplifies to the following partial differential equation (pde) of 3rd order
\begin{align}
\label{EC-GE2}
(1-\ell^2\Delta)\tau_{ij,j}=0\,,
\end{align}
and, using Eq.~(\ref{HL2}), Eq.~(\ref{EC-GE2}) reads 
in terms of the elastic distortion tensor
\begin{align}
\label{EC-dist}
(1-\ell^2\Delta) C_{ijkl}\beta_{kl,j}=0\,.
\end{align}
Following \citet{Jaunzemis}, the polarization of the Cauchy-like stress,
sometimes called `total stress tensor', is defined by
\begin{align}
\label{stress-total}
\sigma_{ij}:=(1-\ell^2\Delta)\tau_{ij}\,.
\end{align}
Then the equilibrium condition~(\ref{EC-GE2}) reads in terms of the 
total stress tensor 
\begin{align}
\label{EC-total}
\sigma_{ij,j}=0\,.
\end{align}

On the other hand, using the so-called `Ru-Aifantis theorem'~\citep{RA}
in terms of stresses, Eq.~(\ref{EC-GE2}) can be written as an equivalent 
system of pdes of 1st order and of 2nd order, namely
\begin{align}
\label{RA1}
\sigma_{ij,j}^0=0\,,\\
\label{RA2}
(1-\ell^2\Delta)\tau_{ij}=\sigma_{ij}^0\,,
\end{align}
where $\sigma^0_{ij}$ is the classical stress tensor.
Eqs.~(\ref{RA1}) and (\ref{RA2}) also play the role of 
the basic equations in Aifantis' version of gradient elasticity
(see, e.g., \citet{AA11}).
Using the `Ru-Aifantis theorem', the total stress tensor
$\sigma_{ij}$ is identified with the classical stress tensor $\sigma_{ij}^0$: 
\begin{align}
\label{T-equiv}
\sigma_{ij}\equiv\sigma_{ij}^0\,.
\end{align}

The so-called `Ru-Aifantis theorem' is a special case of a more
general technique well-known in the theory of partial differential equations  
(see, e.g.,~\citet{Vekua}).
Moreover,  the `Ru-Aifantis theorem' is restricted only to situations involving a
body of infinite extent (with no need to enforce boundary conditions). 
In the presence of boundary conditions, the `Ru-Aifantis theorem' 
is no longer valid and can lead to erroneous solutions.
Therefore, it is questionable if the `Ru-Aifantis theorem'  should be used 
in the construction of crack solutions in gradient elasticity.
In physics, such a method of the reduction of the order of higher order field 
equations is known and used in the so-called 
Bopp-Podolsky theory~\citep{Bopp,Podolsky}, which is the 
gradient theory of electrodynamics. In the Bopp-Podolsky theory,
the (linear) field equation is of fourth order and can be decomposed into 
two partial differential equations of second order (e.g.,~\citet{Davis}).

Now, comparing the theory of `weakly' nonlocal elasticity and 
the theory of gradient elasticity of Helmholtz type one can say the following:
\begin{itemize}
\item[(i)]
The `weakly' nonlocal theory of elasticity is described by 
the equations~(\ref{EC-NL}), (\ref{EC-cl}) and (\ref{t-HE}), while the gradient
elastic one by the equation~(\ref{EC-GE2}) and considering the 
 `Ru-Aifantis theorem' by the equations~(\ref{RA1}) and (\ref{RA2}).
\item[(ii)]
The equations~(\ref{t-HE}) and  (\ref{RA2}) indicate that the nonlocal stresses 
$t_{ij}$ and the Cauchy-like stresses of gradient elasticity theory $\tau_{ij}$ 
are identical, i.e.
\begin{align}
\label{cond1}
t_{ij}\equiv\tau_{ij}\,,
\end{align}
when the `Ru-Aifantis theorem' is used.
\item[(iii)]
The nonlocal stress tensor $t_{ij}$ is different to the total stress tensor 
$\sigma_{ij}$, as it is evident from (\ref{t-HE}) and (\ref{stress-total}).
\item[(iv)]
The natural boundary conditions in nonlocal elasticity theory are referred to
prescribed values of tractions taken from 
the nonlocal stresses $t_{ij}$, as it is illustrated in Eq.~(\ref{BC-NE}).
In other words the natural boundary conditions are as simple as in the
classical case. On the contrary, in gradient elasticity theory natural boundary
conditions are complicated expressions of double stresses $\tau_{ijk}$, which
in the Helmholtz version of the theory are expressed in terms of the
Cauchy-like stresses $\tau_{ij}$ according to the relation~(\ref{CE-GE}).
More details can be found in~\citet{Polyzos}. 
Aifantis and co-workers, adopting the relation~(\ref{cond1}) 
and ignoring the double stresses in (\ref{BC-GE})
have considered that natural boundary conditions in gradient elasticity are similar to
nonlocal ones, i.e.
\begin{align}
\label{BC-NE-A}
\tau_{ij}n_j=\hat{\tau}_i\,.
\end{align}
However, such an assumption is arbitrary since it cannot be supported by 
a variational consideration~(\ref{BC-GE}).
\end{itemize}

In addition, in gradient elasticity of Helmholtz type the following
inhomogeneous Helmholtz equation can be found for the elastic distortion 
$\beta_{ij}$ (e.g.,~\citet{LM05,LM06,Lazar13})
\begin{align}
\label{B-HE}
(1-\ell^2\Delta)\beta_{ij}=\beta^0_{ij},
\end{align}
where $\beta^0_{ij}$ denotes the classical elastic distortion tensor.

\section{Non-singular mode III crack solutions provided by \citet{Aifantis09,Aifantis12}}

The classical solution of the stress produced by a crack of mode III 
obtained in classical fracture mechanics is of the form
\begin{align}
\label{T-zx-0}
\sigma^0_{zx}&=-\frac{K_{\text{III}}}{\sqrt{2\pi r}}\, \sin\frac{\theta}{2}\,,\\
\label{T-zy-0}
\sigma^0_{zy}&=\frac{K_{\text{III}}}{\sqrt{2\pi r}}\, \cos\frac{\theta}{2}\,,
\end{align}
where $K_{\text{III}}$ is the mode III stress intensity factor, and $r=\sqrt{x^2+y^2}$, $\theta=\arctan y/x$ are the polar coordinates centered at the crack tip. 
The classical elastic distortion tensor of a mode III crack reads
\begin{align}
\label{B-zx-0}
\beta^0_{zx}&=-\frac{K_{\text{III}}}{\mu \sqrt{2\pi r}}\, \sin\frac{\theta}{2}\,,\\\label{B-zy-0}
\beta^0_{zy}&=\frac{K_{\text{III}}}{\mu \sqrt{2\pi r}}\, \cos\frac{\theta}{2}\,.\end{align}
The fields~(\ref{T-zx-0})--(\ref{B-zy-0}) possess a $1/\sqrt{r}$-singularity
at the crack tip,
while the stresses~(\ref{T-zx-0}) and (\ref{T-zy-0}) 
fulfill the equilibrium condition
\begin{align}
\label{}
\sigma^0_{zx,x}+\sigma^0_{zy,y}=0
\end{align}
and the elastic distortions~(\ref{B-zx-0}) and (\ref{B-zy-0}) 
satisfy the compatibility condition
\begin{align}
\label{}
\beta^0_{zy,x}-\beta^0_{zx,y}=0\,.
\end{align}
Moreover, on the free of stresses surface of the crack ($\theta=\pi$) the
boundary condition $\sigma^0_{zy}(r,\pi)=0$  is fulfilled.

Substituting Eqs.~(\ref{T-zx-0}) and (\ref{T-zy-0}) into the inhomogeneous
Helmholtz equation~(\ref{t-HE}) or (\ref{RA2}), the non-singular stress is obtained as
\begin{align}
\label{T-zx}
t_{zx}=\tau_{zx}&=-\frac{K_{\text{III}}}{\sqrt{2\pi}}\, \sin\frac{\theta}{2}\, f_1(r)\,,
\\
\label{T-zy}
t_{zy}=\tau_{zy}&=\frac{K_{\text{III}}}{\sqrt{2\pi}}\, \cos\frac{\theta}{2}\,f_1(r)\,,
\end{align}
where $f_1(r)$ is given by (see Eq.~(\ref{f1}))
\begin{align}
\label{f1-2}
f_{1}(r)=\frac{1}{\sqrt{r}}\, \Big(1-\e^{-r/\ell}\Big)\,.
\end{align}
By construction, Eqs.~(\ref{T-zx}) and (\ref{T-zy}) 
satisfy Eq.~(\ref{EC-GE2}).
In nonlocal elasticity, on the free of stresses surface of the crack ($\theta=\pi$) the
boundary condition $t_{zy}(r,\pi)=0$  is satisfied.

In the same way, substituting Eqs.~(\ref{B-zx-0}) and (\ref{B-zy-0}) 
into the inhomogeneous Helmholtz equation~(\ref{B-HE}), 
the non-singular elastic distortion is found as
\begin{align}
\label{B-zx}
\beta_{zx}&=-\frac{K_{\text{III}}}{\mu\sqrt{2\pi}}\, \sin\frac{\theta}{2}\,f_1(r)\,,
\\
\label{B-zy}
\beta_{zy}&=\frac{K_{\text{III}}}{\mu\sqrt{2\pi}}\, \cos\frac{\theta}{2}\, f_1(r)\,.
\end{align}

Eqs.~(\ref{T-zx})--(\ref{B-zy}) have been first 
published\footnote{It is noted that also~\citet{Lazar03} found these non-singular crack fields of mode III.
However, \citet{Lazar03} did not publish his results.}
by~\citet{Aifantis09,Aifantis12}. 
\citet{Aifantis09,Aifantis12} has adopted the regularization conditions
at $r=0$ and $r=\infty$ proposed by~\citet{Eringen83}.
It is obvious that these fields are
non-singular and zero at the crack tip. 
Extremum stress and distortion occur near the crack tip.

However, although the stresses and elastic distortions provided 
by~(\ref{T-zx})--(\ref{B-zy}) are non-singular at the crack tip, they cannot
be considered as a solution of the nonlocal and gradient mode III fracture mechanics 
problem. 
Eqs.~(\ref{T-zx}) and (\ref{T-zy}) are not a solution for a mode III crack in 
nonlocal elasticity since
they do not satisfy the equilibrium condition~(\ref{EC-NL}): 
\begin{itemize}
\item[(i)]
Indeed, inserting  Eqs.~(\ref{T-zx}) and (\ref{T-zy}) into (\ref{EC-NL})
one obtains instead of zero the nonzero line force at the crack tip
\begin{align}
\label{}
F_z=t_{zx,x}+t_{zy,y}
=\frac{K_{\text{III}}}{\sqrt{2\pi r}}\, \sin\frac{\theta}{2}\,
\frac{\e^{-r/\ell}}{\ell}\,,
\end{align}
which, 
in the limit to classical elasticity this line force becomes zero, i.e.
\begin{align}
\label{}
\lim_{\ell\to 0} F_z =0\,.
\end{align}
\end{itemize}
Also Eqs.~(\ref{T-zx})--(\ref{B-zy}) cannot be considered as solution of the
gradient elastic mode III fracture mechanics problem for the following reasons:
\begin{itemize}
\item[(i)]
The corresponding elastic distortions given by Eqs.~(\ref{B-zx}) and (\ref{B-zy})
do not satisfy the compatibility condition 
of gradient elasticity: 
$\beta_{zy,x}-\beta_{zx,y}=0$. 
Indeed, a nonzero dislocation density of screw dislocations
$\alpha_{zz}$ appears at the crack tip
\begin{align}
\label{Azz}
\alpha_{zz}=\beta_{zy,x}-\beta_{zx,y}
=\frac{K_{\text{III}}}{\mu \sqrt{2\pi r}}\, \cos\frac{\theta}{2}\,
\frac{\e^{-r/\ell}}{\ell}\,,
\end{align}
which, in the limit to classical elasticity, this dislocation density becomes 
zero
\begin{align}
\label{}
\lim_{\ell\to 0} \alpha_{zz} =0\,.
\end{align}
\item[(ii)]
It is easy to see that 
due to Eq.~(\ref{Azz}), the elastic distortions $\beta_{zx}$ and
 $\beta_{zx}$ are not anymore a displacement gradient.
In other words, there is not a displacement field  that supports 
Eqs.~(\ref{B-zx}) and (\ref{B-zy}).
\item[(iii)]
Eqs.~(\ref{T-zx}) and (\ref{T-zy}) describe 
the Cauchy-like stresses $\tau_{ij}$ and not the total stresses $\sigma_{ij}$.
Not using the `Ru-Aifantis theorem', \citet{G03} obtained that near the tip
of a gradient elastic mode III crack, the total stresses are more singular than 
in the classical case appearing a singularity of order $r^{-3/2}$
and, therefore, $\sigma_{ij}\neq\sigma_{ij}^0$.
\end{itemize}

Thus, the conclusion here is that the stresses provided 
by Eqs.~(\ref{T-zx}) and (\ref{T-zy}) 
and the elastic distortion~(\ref{B-zx}) and (\ref{B-zy})
are just formal solutions of the 
inhomogeneous Helmholtz equations~(\ref{RA2}) and (\ref{B-HE})
neglecting the equilibrium and compatibility conditions of the nonlocal and gradient
boundary value problems, respectively.
They produce line forces at the crack tip in nonlocal elasticity.
In addition, since the compatibility condition is not fulfilled in
gradient elasticity 
they produce a certain distribution of screw dislocations at the crack tip.

\section{Non-singular mode I crack solutions provided 
by \citet{Aifantis12,Isaksson13} and \citet{Isaksson12}}

For classical mode I cracks, the stresses are singular at the crack tip
and have the form 
\begin{align}
\label{T-xx-0}
\sigma^0_{xx}&=\frac{K_{\text{I}}}{\sqrt{2\pi r}}\, 
\bigg[\frac{3}{4}\, \cos\frac{\theta}{2}
+\frac{1}{4}\, \cos\frac{5\theta}{2}\bigg]\,,
\\
\label{T-yy-0}
\sigma^0_{yy}&=\frac{K_{\text{I}}}{\sqrt{2\pi r}}\, 
\bigg[\frac{5}{4}\, \cos\frac{\theta}{2}
-\frac{1}{4}\, \cos\frac{5\theta}{2}\bigg]\,,
\\
\label{T-xy-0}
\sigma^0_{xy}&=\frac{K_{\text{I}}}{\sqrt{2\pi r}}\, 
\bigg[-\frac{1}{4}\, \sin\frac{\theta}{2}
+\frac{1}{4}\, \sin\frac{5\theta}{2}\bigg]\,,
\end{align}
where $K_{\text{I}}$ denotes the mode I stress intensity factor.
The fields~(\ref{T-xx-0})--(\ref{T-xy-0}) possess a $1/\sqrt{r}$-singularity.
Eqs.~(\ref{T-xx-0})--(\ref{T-xy-0}) satisfy the equilibrium conditions
\begin{align}
\label{}
\sigma^0_{xx,x}+\sigma^0_{xy,y}&=0\,,\\
\sigma^0_{yx,x}+\sigma^0_{yy,y}&=0\,,
\end{align}
the boundary conditions
\begin{align}
\label{}
\sigma^0_{xy}(r,\pi)&=0\,,\\
\sigma^0_{yy}(r,\pi)&=0\,
\end{align}
and the stress compatibility condition
\begin{align}
\label{CC-0}
\Delta \sigma^0_{ij}
+\frac{1}{1+\nu}\, \big(\sigma^0_{kk,ij} -\delta_{ij}\Delta\sigma^0_{kk}\big)=0\,.
\end{align}

Substituting Eqs.~(\ref{T-xx-0})--(\ref{T-xy-0}) into Eq.~(\ref{t-HE}) or Eq.~(\ref{RA2}),
non-singular fields are obtained and they have the form 
(see also Appendix A) 
\begin{align}
\label{T-xx}
t_{xx}=\tau_{xx}&=\frac{K_{\text{I}}}{\sqrt{2\pi}}\, 
\bigg[\frac{3}{4}\, \cos\frac{\theta}{2}\, f_1(r)
+\frac{1}{4}\, \cos\frac{5\theta}{2}\, f_5(r) \bigg]\,,
\\
\label{T-yy}
t_{yy}=\tau_{yy}&=\frac{K_{\text{I}}}{\sqrt{2\pi}}\, 
\bigg[\frac{5}{4}\, \cos\frac{\theta}{2}\, f_1(r)
-\frac{1}{4}\, \cos\frac{5\theta}{2}\, f_5(r)\bigg]\,,
\\
\label{T-xy}
t_{xy}=\tau_{xy}&=\frac{K_{\text{I}}}{\sqrt{2\pi}}\, 
\bigg[-\frac{1}{4}\, \sin\frac{\theta}{2}\, f_1(r)
+\frac{1}{4}\, \sin\frac{5\theta}{2}\, f_5(r)\bigg]\,,
\end{align}
where again $f_1(r)$ is given by Eqs.~(\ref{f1-2}) and (\ref{f1}), 
and $f_5(r)$ reads (see Eq.~ (\ref{f5})):
\begin{align}
\label{f5-2}
f_{5}(r)=\frac{1}{\sqrt{r}}\, \bigg(1-\frac{6\ell^2}{r^2}+
2\bigg(1+\frac{3\ell}{r}+\frac{3\ell^2}{r^2}\bigg) \e^{-r/\ell}\bigg)\,.
\end{align}
The fields~(\ref{T-xx})--(\ref{T-xy}) are non-singular and zero at the crack tip.

The non-singular stresses\footnote{Also \citet{Lazar10} found 
these non-singular solutions. 
Due to the properties as discussed in this section, he did not 
publish the result.}~(\ref{T-xx})--(\ref{T-xy}) were first published 
by~\citet{Isaksson12} and \citet{Isaksson13}. 
The components~(\ref{T-xx}) and (\ref{T-yy}) were also published 
by~\citet{Aifantis12}. 
In finding non-singular crack fields~\citet{Aifantis12,Isaksson12} and \citet{Isaksson13}
used regularity conditions at $r=0$ and $r=\infty$ (see Appendix A).
\citet{Isaksson13} and \citet{Isaksson12}
have claimed that they found the stresses in nonlocal elasticity of Helmholtz type
and~\citet{Aifantis12} has claimed that he found the stresses in gradient elasticity
of Helmholtz type. 
However, the non-singular
stresses~(\ref{T-xx})--(\ref{T-xy}) do not satisfy the equilibrium
condition~(\ref{EC-NL}).
In nonlocal elasticity, they produce line forces at the crack tip
\begin{align}
\label{Fx}
F_x=t_{xx,x}+t_{xy,y}
=\frac{K_{\text{I}}}{\sqrt{2\pi r}}\, 
\frac{1}{4}\,\cos\frac{\theta}{2}\,\frac{\e^{-r/\ell}}{\ell}\,,\\
\label{Fy}
F_y=t_{yx,x}+t_{yy,y}
=\frac{K_{\text{I}}}{\sqrt{2\pi r}}\, 
\frac{3}{4}\,\sin\frac{\theta}{2}\,\frac{\e^{-r/\ell}}{\ell}\,.
\end{align}
In the limit to classical elasticity these line forces become zero:
\begin{align}
\label{}
\lim_{\ell\to 0} F_x =0\,, \\
\lim_{\ell\to 0} F_y =0\,.
\end{align}

In nonlocal elasticity,
 the boundary condition $t_{xy}(r,\pi)=0$ is not satisfied.
Indeed, from Eq.~(\ref{T-xy}) one obtains
\begin{align}
\label{BC-I}
t_{xy}(r,\pi)=\frac{K_{\text{I}}}{\sqrt{2\pi r}}\,
\bigg[-\frac{1}{4}\, f_1(r)+\frac{1}{4}\, f_5(r)\bigg]
\neq 0\,.
\end{align}

Now we have to investigate if the stresses~(\ref{T-xx})--(\ref{T-xy}) 
possess geometric incompatibilities caused by a distribution of dislocations. 
In order to analyze the incompatibility condition of gradient elasticity 
in terms of stresses, we start from the so-called 
incompatibility tensor $\eta_{ij}$
which is defined in terms of the elastic strain tensor~\citep{Kroener,Kroener81,Teodosiu} 
\begin{align}
\label{ink}
\eta_{ij}=-\epsilon_{ikl}\epsilon_{jmn} e_{ln,km}\,,
\end{align}
where $\epsilon_{ikl}$ is the Levi-Civita tensor.
Taking into account $\tau_{ij,j}=F_i$ and the inverse of the Hooke law,
we may rewrite~(\ref{ink}) as (see, e.g.,~\citet{Kroener,Teodosiu})
\begin{align}
\label{BM-CC}
\Delta \tau_{ij}
+\frac{1}{1+\nu}\, \big(\tau_{kk,ij} -\delta_{ij}\Delta\tau_{kk}\big)
-F_{i,j}-F_{j,i}+\delta_{ij} F_{k,k}=2\mu\,\eta_{ij}\,.
\end{align}
If the incompatibility tensor is zero, Eq.~(\ref{BM-CC}) reduces
to the Beltrami-Michell stress compatibility condition.
On the other hand, 
the incompatibility tensor $\eta_{ij}$ can be given~\citep{Kroener,Kroener81,Teodosiu} 
\begin{align}
\label{eta}
\eta_{ij}=-\frac{1}{2}\,
\big(\epsilon_{ikl}\alpha_{lj,k}+\epsilon_{jkl}\alpha_{li,k}\big)
\end{align}
in terms of the dislocation density tensor
\begin{align}
\label{A-el}
\alpha_{ij}=\epsilon_{jkl}\beta_{il,k}\,.
\end{align}
It states that if $\eta_{ij}$ is nonzero, then
the elastic strain is incompatible due to
dislocations.
On the other hand, if the dislocation density is nonzero, 
the plastic distortion tensor $\beta_{ij}^P$ is nonvanishing:
\begin{align}
\label{A-pl}
\alpha_{ij}=-\epsilon_{jkl}\beta^P_{il,k}\,.
\end{align}
The physical reason is that a dislocation is the elementary carrier of plasticity~\citep{Kroener}.

For plane strain,
only the component $\eta_{zz}$ is nonzero, namely
\begin{align}
\label{eta-zz2}
\eta_{zz}=\alpha_{xz,y}-\alpha_{yz,x}\,.
\end{align}
Using $\tau_{zz}=\nu(\tau_{xx}+\tau_{yy})$, Eq.~(\ref{BM-CC}) simplifies to
\begin{align}
\label{CC3}
-(1-\nu)\Delta \big(\tau_{xx}+\tau_{yy}\big)+F_{k,k}=2\mu\,\eta_{zz}
\,.
\end{align}
From Eqs.~(\ref{T-xx}) and (\ref{T-yy}) we obtain
\begin{align}
\label{T-xx+yy}
\tau_{xx}+\tau_{yy}=\frac{2K_{\text{I}}}{\sqrt{2\pi}}\, 
\cos\frac{\theta}{2}\, f_1(r)\,.
\end{align}
Using Eqs.~(\ref{T-xx+yy}) and (\ref{PDE}), we find
\begin{align}
\label{CC4}
\Delta \big(\tau_{xx}+\tau_{yy}\big)=- \frac{2 K_{\text{I}}}{\sqrt{2\pi r}}\,
\cos\frac{\theta}{2}\,\frac{\e^{-r/\ell}}{\ell^2}\,.
\end{align}
In addition, $\text{div} \BF$ is calculated for Eqs.~(\ref{Fx}) and (\ref{Fy})
as
\begin{align}
\label{CC5}
F_{k,k}=- \frac{1}{2}\,\frac{K_{\text{I}}}{\sqrt{2\pi r}}\,
\cos\frac{\theta}{2}\,\frac{\e^{-r/\ell}}{\ell^2}
+ \frac{1}{4}\, \frac{K_{\text{I}}}{\sqrt{2\pi r}}\,
\cos\frac{3\theta}{2}\,\bigg[1+\frac{\ell}{r}\bigg]
\frac{\e^{-r/\ell}}{\ell^2}
\,.
\end{align}
Substituting Eqs.~(\ref{CC4}) and (\ref{CC5}) into (\ref{CC3}), 
the geometric incompatibility is obtained as
\begin{align}
\label{e-zz}
\eta_{zz}=\frac{1}{2\mu}\,\frac{K_{\text{I}}}{\sqrt{2\pi r}}
\bigg(\frac{3-4\nu}{2}\,
\cos\frac{\theta}{2}
+ \frac{1}{4}\, \cos\frac{3\theta}{2}\,\bigg[1+\frac{\ell}{r}\bigg]\bigg)
\frac{\e^{-r/\ell}}{\ell^2}
\,,
\end{align}
which is caused by a nonzero distribution of edge dislocations,
 $\alpha_{xz}$ and $\alpha_{yz}$,  at the crack tip, namely
\begin{align}
\label{A-edge}
\alpha_{xz}=\beta_{xy,x}-\beta_{xx,y}\,,\qquad
\alpha_{yz}=\beta_{yy,x}-\beta_{yx,y}\,.
\end{align}
Thus, the mode I stresses~(\ref{T-xx})--(\ref{T-xy}) produce a nonzero
compatibility condition in gradient elasticity 
in contrast to the classical mode I stresses 
(\ref{T-xx-0})--(\ref{T-xy-0}), which give zero for the incompatibility
and the dislocation density.
In the limit to classical elasticity the geometric incompatibility~(\ref{e-zz})
becomes zero:
\begin{align}
\label{}
&\lim_{\ell\to 0} \eta_{zz} =0\,.
\end{align}

Consequently, the non-singular stresses~(\ref{T-xx})--(\ref{T-xy}) cannot be
considered as a solution of the nonlocal elastic mode I fracture mechanics
problem for the following reasons:
\begin{itemize}
\item[(i)]
They do not fulfill the equilibrium condition of nonlocal elasticity~(\ref{EC-NL}).
\item[(ii)]
They are not compatible with a free of stresses crack, as it is evident by
relation~(\ref{BC-I}).
\end{itemize}
Also, Eqs.~(\ref{T-xx})--(\ref{T-xy}) do not represent the correct stresses 
of a gradient elastic mode I fracture mechanics problem, because:
\begin{itemize}
\item[(i)] 
They do not satisfy the compatibility conditions of gradient theory of compatible
elasticity, as it is evident from Eqs.~(\ref{BM-CC}) and (\ref{e-zz}). 
\item[(ii)]
As in the case of mode III crack, the elastic distortions 
$\beta_{xx}$, $\beta_{xy}$, $\beta_{yx}$ and $\beta_{yy}$ cannot be taken as a
gradient of a displacement vector.
\item[(iii)] 
As it has been mentioned in the case of mode III crack, 
Eqs.~(\ref{T-xx})--(\ref{T-xy}) represent the Cauchy-like stresses $\tau_{ij}$
and not the total stresses $\sigma_{ij}$. 
As it is shown by~\citet{Karlis} and \citet{GG09},  
ignoring the `Ru-Aifantis theorem', 
total stresses are more singular than in the classical case appearing a
singularity of order $r^{-3/2}$ near the tip of a gradient elastic mode I
crack and, therefore, $\sigma_{ij}\neq\sigma_{ij}^0$.
\end{itemize}

Thus, the conclusion is that the non-singular stress 
components~(\ref{T-xx})--(\ref{T-xy})
are formal solutions of an inhomogeneous Helmholtz equation.
They produce line forces in nonlocal elasticity and a distribution of edge dislocations 
at the crack tip in gradient elasticity.

\section{Discussion}

The main conclusion of the present work is that the non-singular stress fields
provided by~\citet{Aifantis09,Aifantis12,Isaksson13} and \citet{Isaksson12}
cannot be considered as the solution of a nonlocal or strain gradient elastic
fracture mechanics problem. 
As it has been already mentioned, in the framework of nonlocal elasticity
Eringen and co-workers~\citep{Eringen02} were the first who proposed
non-singular stress fields near the mode I and mode III crack tips by solving
rigorously the corresponding nonlocal fracture mechanics problem.

On the other hand, 
in classical fracture mechanics and in compatible gradient elasticity 
(e.g.,~\citet{G03,GG09}), the mathematical solutions of cracks that take into account
the equations of elasticity, the equilibrium equations, boundary conditions,
and compatibility conditions lead to nice, exact description of the high
magnitude crack tip stress field (classical singularity $r^{-1/2}$ and 
non-classical singularity $r^{-3/2}$). 
\citet{Karlis} solving numerically the strain gradient elastic mode I crack
problem provided plots for the variation of $r^{-1/2}$- and  $r^{-3/2}$-terms
as a function of the internal length scale parameter $\ell$.

However, such solutions do not answer the 
question: Why is the stress at the crack tip so big?
\citet{Weertman} gave a physical answer that the stress is caused by a 
distribution of many dislocations near the crack tip. 
Also \citet{Weertman} pointed out that fracture mechanics cannot be understood
at a deeper level unless its study includes dislocations.
In this paper, we found that the non-singular crack solutions 
do not fulfill the compatibility conditions. That means that dislocations are
behind these non-singular crack solutions and it does not make sense anymore to
require to fulfill a compatibility condition. 
A dislocation is the building block of a crack as mentioned by \citet{Weertman}.
Unlike a crack, a dislocation is an elementary defect in solids able to build 
composite defects.
For that reason the incompatibility of a crack is expressed in terms of
dislocation densities.

Indeed for a mode III crack, 
due to Eq.~(\ref{Azz}), the elastic distortions $\beta_{zx}$ and
 $\beta_{zx}$ are not anymore a displacement gradient. In fact,
the elastic distortions are incompatible and the incompatible parts are the 
plastic distortions $\beta_{zx}^P$ and $\beta_{zy}^P$:
\begin{align}
\label{beta-screw}
\beta_{zx}=u_{z,x} -\beta_{zx}^P\,,\qquad
\beta_{zy}=u_{z,y} -\beta_{zy}^P\,,
\end{align}
and there is not a displacement field $u_z$ which provides the elastic 
distortions~(\ref{B-zx}) and (\ref{B-zy}) as compatible displacement gradient.
In the framework of compatible elasticity, 
Eqs.~(\ref{B-zx}) and (\ref{B-zy}) cannot be considered as a solution
of the gradient elastic mode III fracture mechanics problem since
they do not satisfy the compatibility condition. 
The only way-out is the interpretation 
of the incompatible elastic distortions~(\ref{B-zx}) and (\ref{B-zy}) 
in the framework of incompatible elasticity or 
a dislocation-based mode III fracture problem (see also~\citet{Weertman93,Weertman}).
In dislocation-based fracture mechanics, a dislocation is the basic building
block of the crack and fracture mechanics can be developed from its dislocation foundation.
Then $\alpha_{zz}$ plays the role of distribution of non-redundant dislocations
which are known also as `geometrically necessary' dislocations. 
The density of non-redundant dislocations may define the `plastic' zone near the crack tip
(see Fig. 1). 
A distribution of non-redundant screw dislocations appears 
within the plastic zone of the crack tip.
$\alpha_{zz}$ is singular at the crack tip and zero at the crack faces.
\begin{figure}[t!!]\unitlength1cm
\centerline{
(a)
\begin{picture}(7,7)
\put(0.0,0.2){\epsfig{file=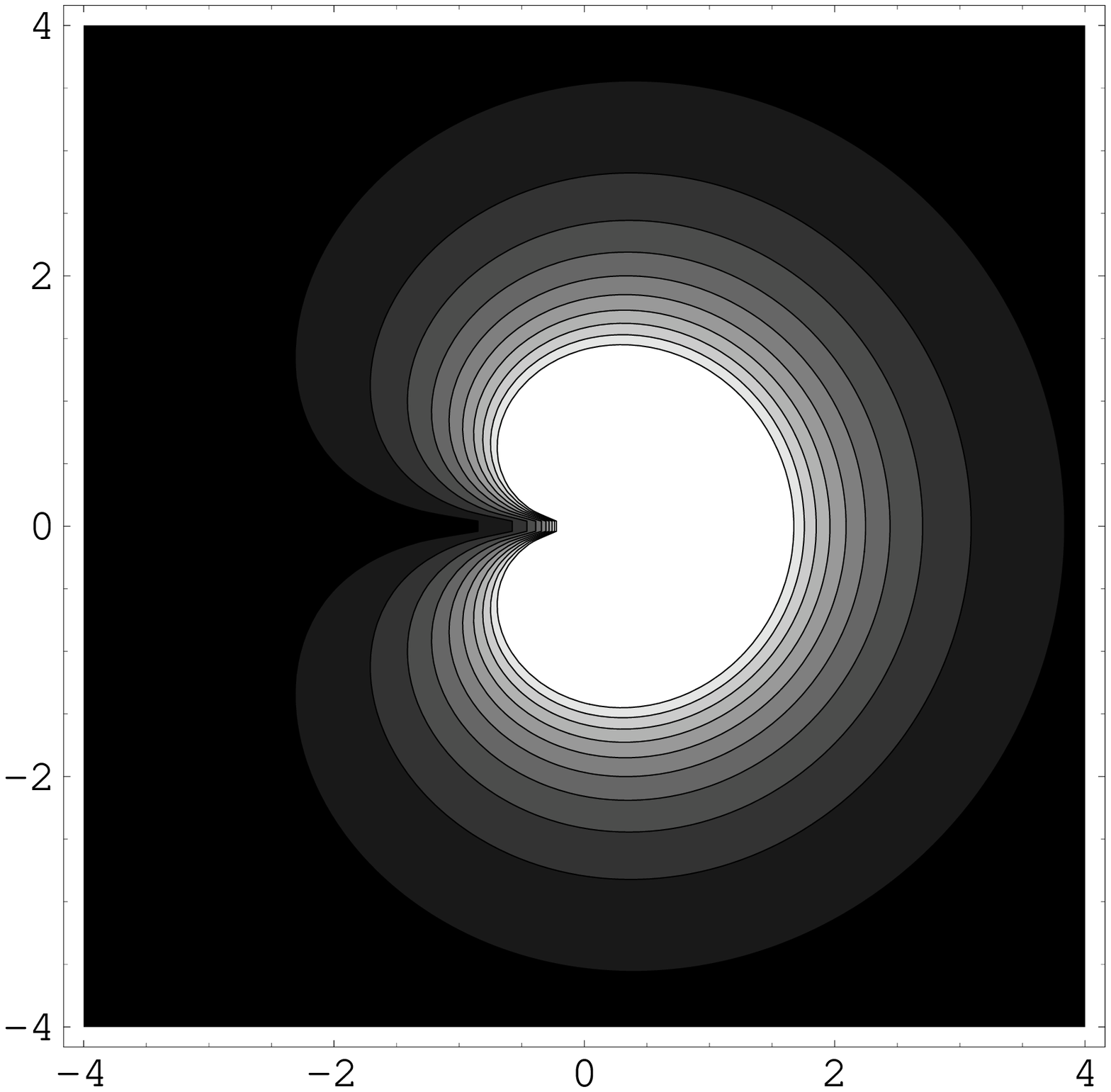,width=6.8cm}}
\put(3.5,-0.1){$x$}
\put(-0.3,3.6){$y$}
\end{picture}
(b)
\begin{picture}(7,7)
\put(0.0,0.2){\epsfig{file=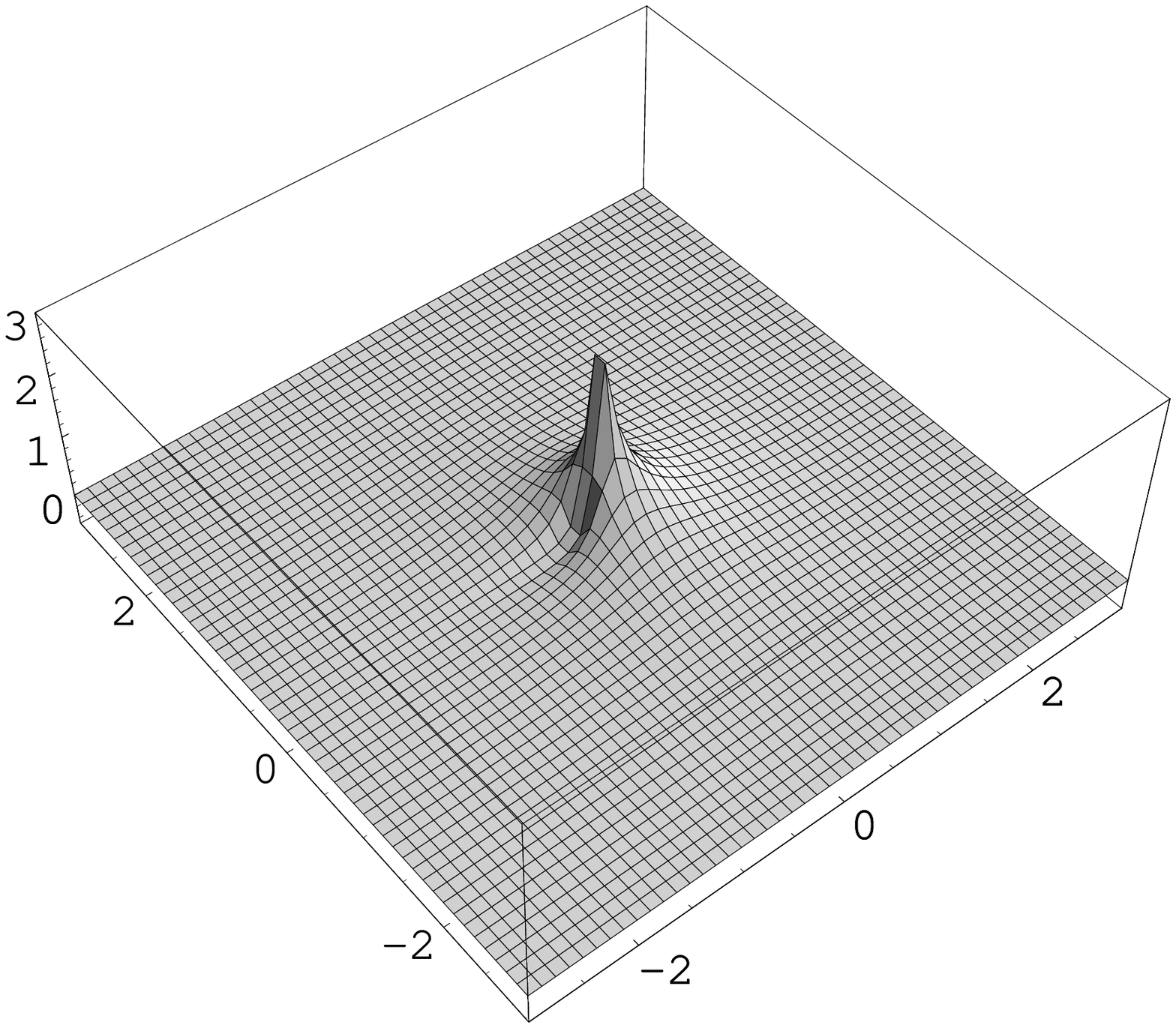,width=7.8cm}}
\put(6.0,0.6){$x$}
\put(1.0,1.3){$y$}
\end{picture}
}
\caption{Dislocation density $\alpha_{zz}$ of a mode III crack:
(a)  contour plot and (b) 3D plot.}
\label{fig:alpha}
\end{figure}

The same can be shown for a mode I  crack.  
Due to nonzero dislocation density of edge dislocations $\alpha_{xz}$ and
$\alpha_{yz}$ and nonzero incompatibility $\eta_{zz}$,
the corresponding elastic distortions are not 
a simple gradient of the displacements $u_x$ and $u_y$.  
They are incompatible due to 
nonzero plastic distortions:
\begin{align}
\label{beta-edge1}
\beta_{xx}=u_{x,x} -\beta_{xx}^P\,,\qquad
\beta_{xy}=u_{x,y} -\beta_{xy}^P\,,\\
\label{beta-edge2}
\beta_{yx}=u_{y,x} -\beta_{yx}^P\,,\qquad
\beta_{yy}=u_{y,y} -\beta_{yy}^P\,.
\end{align}
As for the mode III crack problem, 
a possible interpretation of the incompatible elastic distortions may be 
a dislocation-based mode I fracture problem (see also~\citet{Weertman93,Weertman}).

\section*{Acknowledgement}
M.L. gratefully acknowledges the grants obtained from the 
Deutsche Forschungsgemeinschaft (Grant Nos. La1974/2-1, La1974/2-2, La1974/3-1).

\begin{appendix}
\section{Appendix}
\setcounter{equation}{0}
\renewcommand{\theequation}{\thesection.\arabic{equation}}

Solving the inhomogeneous Helmholtz equations for cracks 
requires finding the solution of the partial differential equation of the form
\begin{align}
\label{PDE}
\big[1-\ell^2\Delta]\, g_n(r,\theta)=\frac{1}{\sqrt{r}}\, \e^{\ii n \theta/2}
\,,\quad n=\pm 1,\pm 5\,,
\end{align}
where the Laplacian reads in polar coordinates
\begin{align}
\Delta=\frac{\pd^2}{\pd r^2}+\frac{1}{r}\, \frac{\pd}{\pd r}
+\frac{1}{r^2}\, \frac{\pd^2}{\pd \theta^2}
\end{align}
and 
\begin{align}
g_n(r,\theta)=f_n(r)\, \e^{\ii n \theta/2}\,,
\end{align}
where $f_n(r)$ is the radial part of the solutions.
Using the regularity conditions at $r=0$ and $r=\infty$, namely $f_n(r)$
should be bounded at $r=0$ and $r=\infty$, we find
\begin{align}
\label{f1}
f_{\pm 1}(r)=\frac{1}{\sqrt{r}}\, \Big(1-\e^{-r/\ell}\Big)
\end{align}
and
\begin{align}
\label{f5}
f_{\pm 5}(r)=\frac{1}{\sqrt{r}}\, \bigg(1-\frac{6\ell^2}{r^2}+
2\bigg(1+\frac{3\ell}{r}+\frac{3\ell^2}{r^2}\bigg) \e^{-r/\ell}\bigg)\,.
\end{align}
It is easy to see that $f_n(r)$ is zero at $r=0$.
The near fields are
\begin{align}
f_{\pm 1}(r)&\approx
\frac{r^{1/2}}{\ell}-\frac{r^{3/2}}{2\ell^2}+\frac{r^{5/2}}{6\ell^3}
-\frac{r^{7/2}}{24\ell^4}+\cdots\,,\\
f_{\pm 5}(r)&\approx
\frac{r^{3/2}}{4\ell^2}-\frac{2 r^{5/2}}{15\ell^3}
+\frac{r^{7/2}}{24\ell^4}-\cdots\,.
\end{align}
It is worth noting that also \citet{Eringen83} derived the solution~(\ref{f1}). 
The expression for $f_{\pm 5}(r)$ was given by \citet{Eringen83} 
in a more sophisticated integral representation.

\end{appendix}

\end{document}